\documentstyle{article}
\def\be{\begin{equation}}
\def\ee{\end{equation}}
\def\bea{\begin{eqnarray}}
\def\eea{\end{eqnarray}}
\def\l{\label}
\def\c{\cite}
\def\r{\ref}
\newcommand{\sfrac}[2]{{\textstyle{#1\over#2}}}
\def\case#1/#2{\textstyle\frac{#1}{#2}}
\def\k0{\kappa_{0}}

\title{\sc Kinematics and Dynamics of $f(R)$ Theories of Gravity}
\vspace{0.6cm}
\normalsize
\vspace{2cm}
\author{{\sc Steve Rippl$^{1}$\thanks{
e-mail: S.F.Rippl@maths.qmw.ac.u},
Henk van Elst$^{1}$\thanks{e-mail: H.van.Elst@maths.qmw.ac.uk},
Reza Tavakol$^{1}$\thanks{e-mail: R.K.Tavakol@maths.qmw.ac.uk} \&
David Taylor$^{2,3}$\thanks{e-mail: TAYLOR@gauss.cam.wits.ac.za}}
\\
\small{$^1$ {\em Astronomy Unit, School of Mathematical Sciences,
Queen Mary \& Westfield College, Mile End Road}}\\
\small{\em London E1 4NS, UK}\\
\\
\small{$^2$ {\em Department of Computational and Applied
Mathematics}}\\
\small{\em University of the Witwatersrand,
PO WITS, 2050, South Africa}\\
\\
\small{$^3$ {\em School of Mathematical Studies,
University of Portsmouth,
PO1 2EG, UK}} }
\date{\normalsize{May 3, 1995}}
\vspace{2cm}
\begin{document}
\sloppy
\maketitle

\begin{abstract}
We generalise the equations governing relativistic fluid dynamics
given by Ehlers and Ellis for general relativity, and by Maartens
and Taylor for quadratic theories, to generalised $f(R)$ theories
of gravity. In view of the usefulness of this alternative framework
to general relativity, its generalisation can be of potential
importance for deriving analogous results to those obtained in
general relativity. We generalise, as an example, the results of
Maartens and Taylor to show that within the framework of general
$f(R)$ theories, a perfect fluid spacetime with vanishing
vorticity, shear and acceleration is
Friedmann--Lema\^{\i}tre--Robertson--Walker only if the fluid has
in addition a barotropic equation of state. It then follows that
the Ehlers--Geren--Sachs theorem and its ``almost'' extension also
hold for $f(R)$ theories of gravity.
\end{abstract}
\vspace{2cm}
\begin{center}
PACS number(s): 04.50.+h, 98.80.Hw, 98.80.Cq
\end{center}
\vfill
\newpage
\section{Introduction}
Recently, Maartens and Taylor \cite{maatay94} examined the
kinematical and dynamical properties of 4-D fluid spacetimes in
quadratic gravity --- with the gravitational Lagrangian density of
the type
\be
{\cal L}_{G} = \sfrac{1}{2\k0}\,\sqrt{-g}\,\left[\,R
+ \alpha\,R^{2} + \beta\,R_{\mu\nu}\,R^{\mu\nu}\,\right]\ ,
\ee
where $\alpha$ and $\beta$ are positive coupling constants, $R$ is
the Ricci curvature scalar and $R_{\mu\nu}$ is the Ricci curvature
tensor. They derived the generalisations of the equations governing
relativistic fluid dynamics given by Ehlers \c{ehl61} and Ellis
\c{ell71} for general relativity (GR) to the quadratic case.
The kinematic and dynamic formulation of fluid spacetimes has been
extremely useful in studying a variety of problems within the
context of GR and it is similarly expected to be of value in the
case of higher-order theories of gravity, as it gives an
alternative framework to study problems analogous to those
considered in GR. An interesting example of this was given by
Maartens and Taylor, when they showed that the usual kinematic
characterisation of the Friedmann--Lema\^{\i}tre--Robertson--Walker
(FLRW) geometry no longer holds in the quadratic theory, but that
it can be restored, if in addition the fluid is assumed to be
barotropic. A consequence of this is that the Ehlers--Geren--Sachs
theorem (EGS) \c{egs68} also holds in the setting of quadratic
theories of gravity.  The EGS theorem states that if a family of
freely falling observers measure the self-gravitating background
radiation to be everywhere exactly isotropic, then the Universe is
exactly homogeneous. The importance of this theorem lies in the
fact that the observable isotropy of the cosmic microwave
background radiation (CMBR) may be employed\footnote{Assuming the
isotropy to hold everywhere.} to deduce that the spacetime is also
homogeneous and therefore of the FLRW type.
\\

A shortcoming of the above work is that, as is customary, it
confines itself to the quadratic theories, thereby implicitly
assuming the quadratic theories to be the generic prototypes of the
higher-order theories in general.  Higher-order corrections to the
action arise in attempts at generalising GR as for example in (i)
the low energy limit of superstrings \c{candela}, (ii) in
perturbation expansions of GR, where the existence of such
higher-order terms can eliminate the divergences that would
otherwise arise \c{antoniadis,barth} and (iii) in the
renormalisation of higher-loop contributions which gives rise to
terms in the action that are higher than quadratic. Such terms are
thought to have been dominant at early epochs in the evolution of
the Universe, when the curvature of spacetime was high.

This indicates that the genericity of the quadratic theories cannot
be assumed a priori and higher than quadratic terms in the action
are also likely to be involved in generalisations of GR. Further,
there is no a priori reason why results obtained within the context
of quadratic theories remain stable (in the sense of \cite{acrt})
with respect to cubic (and higher-order) perturbations to the
action. In fact there are examples where the inclusion of cubic
terms to the action can produce qualitatively important changes to
the corresponding results obtained within the framework of the
quadratic theories (see \c{hvejlrt} for an example). To have a feel
for this, recall that general higher-order theories with
gravitational Lagrangian density of the form
\be
\l{action}
{\cal L}_{G} = \sfrac{1}{2\k0}\,\sqrt{-g}\,f(R)\ ,
\l{fr}
\ee
where $f(R)$ is a differentiable function of $R$ and $df(R)/dR \neq
0$, can be expressed as GR plus a scalar field $\phi$, with a
corresponding scalar potential $V(\phi)$ \c{barcos,maeda}. Now it
turns out that the asymptotic behaviour of this potential depends
crucially on the form of the action. For example, in 4-D
spacetimes, the scalar potential $V(\phi)$ corresponding to
quadratic theories tends asymptotically towards a finite constant
limiting value, whereas for cubic {\it and} higher-order actions,
it tends to zero \c{barcos}. Higher than quadratic terms in the
action could therefore produce important consequences.

As a result, it is of importance to generalise the equations
governing relativistic fluid dynamics given by Maartens and Taylor
to the case of general higher-order gravity with gravitational
Lagrangian density given by (\r{fr}). This would be of value, as it
allows an alternative framework within which general higher-order
theories can be studied.  Here as an example, we study the
conditions for the kinematic characterisation of the FLRW models
(and the EGS theorem), to see whether they are stable
with respect to this action. We then use our results to test the
stability of the almost-EGS theorem of \c{sme95,tm95}.
\\

The plan of the paper is as follows. In section 2 we generalise the
results of \c{maatay94}, by deriving the Ehlers--Ellis equations of
relativistic fluid dynamics for a general action with a
gravitational Lagrangian density of the form (\r{action}) and a
general energy-momentum tensor $T_{\mu\nu}$. In section 3 we
consider the FLRW models, the EGS and almost-EGS theorems
within the context of these theories, and finally section 4
contains the conclusion.
\section{Fluid Kinematics and Dynamics of $f(R)$ Theories}
In this section we generalise the equations of relativistic fluid
dynamics given in \c{ehl61,ell71} to the case of generalised
theories of gravity given by the action
\be
\l{act}
{\cal S} = \frac{1}{2\k0}\ \int_{\cal M}\ \sqrt{-g}\ f(R)\
d^{4}x + \int_{\cal M}\ {\cal L}_{M}\ d^{4}x\ ,
\ee
where $f(R)$ is some arbitrary differentiable function of the Ricci
curvature scalar, ${\cal L}_{M}$ is the Lagrangian density for the
matter sources and we have chosen units such that $c=1$ and
$\k0=8\pi\,G$.  Throughout, Greek indices run from
$\mu,\,\nu=0\dots 3$ and in a coordinate system $\{x^0, x^{a}\}~ (a
=1,2,3)$, the spatial coordinates are denoted by Latin indices
$a,b, \dots$.
\\

The variation of this action with respect to the metric tensor of
spacetime, $g_{\mu\nu}$, results in the generalised field
equations, given by \c{buch70}
\bea
\l{ric}
R_{\mu\nu} & = & f'\,^{-1}\ \left[\ \sfrac{1}{2}\,f\,g_{\mu\nu}
+ f''\,\left(\nabla_{\mu}\nabla_{\nu}R - g_{\mu\nu}\,\Box R\right)
\right.\nonumber\\
& & + \ \left.f'''\,\left(\nabla_{\mu}R\,\nabla_{\nu}R-g_{\mu\nu}\,
\nabla_{\rho}R\,\nabla^{\rho}R\right)
+ \k0\,T_{\mu\nu}\ \right]
\ ,
\eea
where $f':=df(R)/dR$, etc, $\Box R:=g^{\mu\nu}\,\nabla_{\mu}
\nabla_{\nu}R$, and the energy-momentum tensor of the matter
sources is defined by the variational derivative
$T^{\mu\nu}:=(2/\sqrt{-g})\,\delta{\cal L}_{M}/\delta g_{\mu\nu}$.
(Note that $\nabla_{\mu}T^{\mu\nu} = 0$ still holds \c{buch70}.)
\\

The contraction of this equation gives the Ricci curvature scalar
in the form
\be
\l{rscl}
R = f'\,^{-1}\ \left[\ 2f - 3f''\,\Box R - 3f'''\,
\nabla_{\mu}R\,\nabla^{\mu}R
+ \k0\,T\ \right]\ ,
\ee
where $T$ denotes the trace of $T_{\mu\nu}$. Taking the covariant
derivative of the Ricci scalar (\r{rscl}) one obtains
\be
\l{id1}
\nabla_{\mu}R = \frac{3f''\,\nabla_{\mu}\left(\Box R\right)
+ 6f'''\,\left(\nabla_{\mu}\nabla_{\nu}R\right)\nabla^{\nu}R
- \k0\,\nabla_{\mu}T}{f'-f''\,R
- 3f'''\,\Box R - 3f''''\,\nabla_{\rho}R\,\nabla^{\rho}R}\ .
\ee

Now following Ehlers \c{ehl61} and Ellis \c{ell71}, we define a
projection tensor, which allows geometrical quantities to be
projected into the instantaneous rest spaces of observers
co-moving with the timelike matter flow with (normalised)
4-velocity $u^{\mu}$, given by $h^{\mu}\!_{\nu}:=
\delta^{\mu}\!_{\nu}+u^{\mu}\,u_{\nu}$.
Projecting parallel and orthogonal to the matter flow lines one
obtains a $1+3$ decomposition of the dynamic equations (``threading
picture''). In particular this yields a decomposition of the
covariant derivative of the matter 4-velocity $u^{\mu}$, which is
used to define kinematical quantities characteristic of the matter
flow, in the form
\be
\nabla_{\nu}u_{\mu} := -\,a_{\mu}\,u_{\nu} + \sigma_{(\mu\nu)}
+ \sfrac{1}{3}\,\Theta\,h_{\mu\nu} + \omega_{[\mu\nu]}\ ,
\ee
where $a^{\mu}$ denotes the acceleration vector,
$\sigma_{(\mu\nu)}$ the rate of shear tensor, $\Theta$ the rate of
expansion scalar, and $\omega_{[\mu\nu]}$ the rate of vorticity
tensor. Also in line with \c{ehl61,ell71}, one can define the usual
vorticity vector in the form ($\epsilon_{0123}=\sqrt{-g}$)
$\omega^{\mu}:=\sfrac{1}{2}\,\epsilon^{\mu\nu\rho\sigma}\,
\omega_{\nu\rho}\,u_{\sigma} \Rightarrow
\omega_{\mu\nu}=\epsilon_{\mu\nu\rho\sigma}\,\omega^{\rho}\,
u^{\sigma}$.
\\

The ``electric'' and ``magnetic parts'' of the trace-free Weyl
conformal curvature tensor can be defined by
\be
E_{(\mu\nu)}(u) := C_{\mu\rho\nu\sigma}\,u^{\rho}\,u^{\sigma},
\hspace{2cm}
H_{(\mu\nu)}(u) := \sfrac{1}{2}\,\epsilon_{\mu\rho\alpha\beta}\,
C^{\alpha\beta}\!_{\nu\sigma}\,u^{\rho}\,u^{\sigma}\ .
\ee
\\

Splitting the field equations as in the case of GR (and
the quadratic case \c{maatay94}), by contracting the Ricci tensor
(\r{ric}) with the 4-velocity $u^{\mu}$ and the projection tensor
$h^{\mu}\!_{\nu}$, yields
\bea
\l{id2}
R_{\mu\nu}\,u^{\mu}\,u^{\nu} & = & f'\,^{-1}\,\left[\ -\sfrac{1}{2}
\,f + f''\,h^{\mu\nu}\,\nabla_{\mu}\nabla_{\nu}R
+ f'''\,h^{\mu\nu}\,\nabla_{\mu}R\,\nabla_{\nu}R
+ \k0\,T_{\mu\nu}\,u^{\mu}\,u^{\nu}\ \right]\\
\l{id3}
R_{\nu\rho}\,h^{\nu}\!_{\mu}\,u^{\rho} & = & f'\,^{-1}\,\left[\
f''\,h^{\nu}\!_{\mu}\left(\nabla_{\nu}\nabla_{\rho}R\right)
u^{\rho} + f'''\,\dot{R}\ h^{\nu}\!_{\mu}\,\nabla_{\nu}R
+ \k0\,T_{\nu\rho}\,h^{\nu}\!_{\mu}\,u^{\rho}\ \right]\\
\l{id4}
R_{\rho\sigma}\,h^{\rho}\!_{\mu}\,h^{\sigma}\!_{\nu} & = &
f'\,^{-1}\,\left[\ \sfrac{1}{2}\,f\,h_{\mu\nu} + f''\,
h^{\rho}\!_{\mu}\,h^{\sigma}\!_{\nu}\left(\nabla_{\rho}
\nabla_{\sigma}R - h_{\rho\sigma}\,\Box R\right)\right.\nonumber\\
& & + \ \left.f'''\,h^{\rho}\!_{\mu}\,h^{\sigma}\!_{\nu}\left(
\nabla_{\rho}R\,\nabla_{\sigma}R - h_{\rho\sigma}\,\nabla_{\lambda}
R\,\nabla^{\lambda}R\right) + \k0\,T_{\rho\sigma}\,
h^{\rho}\!_{\mu}\,h^{\sigma}\!_{\nu}\ \right].
\eea
As in the case of GR \c{ehl61,ell71}, starting from the Ricci
identities for the vector field $u^{\mu}$, multiplying by $u^{\nu}$
and projecting into the local 3-space, gives the propagation
equation for $v_{\mu\nu} = h^{\alpha}\!_{\mu}\,h^{\beta}\!_{\nu}\,
\nabla_{\alpha}u_{\beta}$, in the form
\be
\l{id5}
h^{\alpha}\!_{\mu}\,h^{\beta}\!_{\nu}\,\dot{v}_{\alpha\beta}
- a_{\mu}\,a_{\nu} - h^{\alpha}\!_{\mu}\,h^{\beta}\!_{\nu}\,
\nabla_{\alpha}a_{\beta} + v^{\rho}\!_{\mu}\,v_{\nu\rho}
+ R_{\alpha\gamma\beta\delta}\,h^{\alpha}\!_{\mu}\,u^{\gamma}\,
h^{\beta}\!_{\nu}\,u^{\delta} = 0\ .
\ee
The trace of this equation yields the generalised Raychaudhuri
equation for the action (\r{act})
\bea
\l{id6}
0 & = & \dot{\Theta} + \sfrac{1}{3}\,\Theta^2 - \nabla_{\mu}a^{\mu}
+ 2\left(\sigma^{2} - \omega^{2}\right)\nonumber\\
& & \ +  \ f'\,^{-1}\,\left[\ -\sfrac{1}{2}\,f
+ f''\,h^{\mu\nu}\,\nabla_{\mu}\nabla_{\nu}R
+ f'''\,h^{\mu\nu}\,\nabla_{\mu}R\,\nabla_{\nu}R
+ \k0\,T_{\mu\nu}\,u^{\mu}\,u^{\nu}\ \right]\ .
\l{constraint}
\eea
The skew part of (\r{id5}) is the generalised vorticity propagation
equation for the $f(R)$ theories,
\be
\l{id7}
0 = h^{\alpha}\!_{[\mu}\,h^{\beta}\!_{\nu]}
\left(\dot{\omega}_{\alpha\beta}+\nabla_{\alpha}a_{\beta}\right)
- 2\,\sigma^{\rho}\!_{[\mu}\,\omega_{\nu]\rho} + \sfrac{2}{3}\,
\Theta\,\omega_{\mu\nu}\ ,
\ee
which is identical to the Einstein case. On the other hand the
symmetric, trace-free part of (\r{id5}) is the shear propagation
equation for the action (\r{act}) given by
\bea
\l{id8}
0 & = & h^{\alpha}\!_{(\mu}\,h^{\beta}\!_{\nu)}\left(
\dot{\sigma}_{\alpha\beta}-\nabla_{\alpha}a_{\beta}\right)
- a_{\mu}\,a_{\nu} + \omega_{\mu}\,\omega_{\nu} +
\sigma_{\mu\rho}\,\sigma^{\rho}\!_{\nu} + \sfrac{2}{3}\,\Theta\,
\sigma_{\mu\nu}\nonumber\\
& & + \,\sfrac{1}{3}\,h_{\mu\nu}\left(\nabla_{\rho}a^{\rho}
- \omega^{2} - 2\sigma^{2}\right) + E_{\mu\nu}\nonumber\\
& & - \,\sfrac{1}{2}\,f'\,^{-1}\left(h^{\alpha}\!_{\mu}\,
h^{\beta}\!_{\nu}-\sfrac{1}{3}\,h_{\mu\nu}\,h^{\alpha\beta}\right)
\left[\ f''\,\nabla_{\alpha}\nabla_{\beta}R + f'''\,\nabla_{\alpha}
R\,\nabla_{\beta}R + \k0\,T_{\alpha\beta}\ \right]\ .
\eea
One can also obtain the three constraint equations from the Ricci
identities for $u^{\mu}$ in the form
\bea
0 & = & h^{\mu}\!_{\nu}\left(\sfrac{2}{3}\,\nabla^{\nu}\Theta
- h^{\rho}\!_{\sigma}\,\nabla_{\rho}\sigma^{\nu\sigma}\right)
+ \epsilon^{\mu\nu\rho\sigma}\,u_{\sigma}\left(\nabla_{\nu}
\omega_{\rho} - 2\,\omega_{\nu}\,a_{\rho}\right)\nonumber\\
& & + \ f'\,^{-1}\,\left[\
f''\,h^{\mu}\!_{\nu}\left(\nabla^{\nu}\nabla_{\rho}R\right)
u^{\rho} + f'''\,\dot{R}\ h^{\mu}\!_{\nu}\,\nabla^{\nu}R
+ \k0\,T^{\nu}\!_{\rho}\,h^{\mu}\!_{\nu}\,u^{\rho}\ \right]\ ,
\l{id9}
\eea
which are the field equations (\r{id3}), together with
\be
\l{id10}
h^{\mu}\!_{\nu}\,\nabla_{\mu}\omega^{\nu} = a_{\mu}\,\omega^{\mu}
\ee
\be
\l{id11}
H_{\mu\nu} = 2\,a_{(\mu}\omega_{\nu)} + h^{\alpha}\!_{(\mu}\,
h^{\beta}\!_{\nu)}\left(\nabla^{\gamma}\omega_{\alpha}\!^{\delta}
+ \nabla^{\gamma}\sigma_{\alpha}\!^{\delta}\right)
\epsilon_{\beta\gamma\delta\epsilon}\,u^{\epsilon}\ .
\ee
\\

Using the decomposition of the Riemann curvature tensor into the
trace-free Weyl conformal curvature tensor, the Ricci curvature
tensor and the Ricci curvature scalar,
\be
R_{\mu\nu\rho\sigma} := C_{\mu\nu\rho\sigma}
+ g_{\mu[\rho}\,R_{\sigma]\nu} + g_{\nu[\sigma}\,R_{\rho]\mu}
- \frac{1}{3}\,R\ g_{\mu[\rho}\,g_{\sigma]\nu}\ ,
\ee
the Bianchi identities
\be
\nabla_{[\mu}R_{\nu\rho]\sigma\tau} = 0
\ee
can be cast into the form \c{kuntru62}
\be
\nabla_{\sigma}C^{\mu\nu\rho\sigma} = - \nabla^{[\mu}R^{\nu]\rho}
- \sfrac{1}{6}\,g^{\rho[\mu}\,\nabla^{\nu]}R\ .
\ee

Now using equations (\r{id1})--(\r{id4}), the constraint and
evolution equations for $E_{\mu\nu}$ and $H_{\mu\nu}$ can be
derived for the case of $f(R)$ theories, in the form
\bea
\l{dive}
h^{\mu}\!_{\alpha}\,h^{\gamma}\!_{\beta}\,\nabla_{\gamma}
E^{\alpha\beta} & = &
\epsilon^{\mu\alpha\beta\gamma}\ \sigma_{\alpha\delta}
\ H^{\delta}\!_{\beta}\ u_{\gamma}
- 3H^{\mu}\!_{\alpha}\,\omega^{\alpha}\nonumber\\
& & + \ \k0\,f'\,^{-1}\,h^{\mu}\!_{\alpha}\,\left[\
h^{\beta}\!_{\gamma}\,\nabla^{[\alpha}T^{\gamma]}\!_{\beta}
- \sfrac{1}{3}\,\nabla^{\alpha}T\ \right]\nonumber\\
& & + \ \sfrac{1}{2}\,f'\,^{-1}\,f''\,
h^{\mu}\!_{\alpha}\,
\left[\ \dot{R}\,R^{\alpha}\!_{\beta}\,u^{\beta}
- \left(R_{\beta\gamma}\,u^{\beta}\,u^{\gamma}
+\sfrac{1}{3}R\right)\nabla^{\alpha}R\right.\nonumber\\
& & - \ \left.\nabla^{\alpha}\left(\Box R\right)
- \left(\nabla^{\alpha}\nabla_{\beta}R\right)
\dot{ }\,u^{\beta} + h^{\beta}\!_{\gamma}\,
\nabla^{\alpha}\left(\nabla_{\beta}\nabla^{\gamma}R\right)
\ \right],\\
\l{edot}
h^{\mu}\!_{\alpha}\,h^{\nu}\!_{\beta}\,\dot{E}^{\alpha\beta} & = &
3E^{(\mu}\!_{\alpha}\,\sigma^{\nu)\alpha} + E^{(\mu}\!_{\alpha}\,
\omega^{\nu)\alpha}
- \Theta\,E^{\mu\nu} - h^{\mu\nu}\,E_{\alpha\beta}\,
\sigma^{\alpha\beta}\nonumber\\
& & - \ 2H^{(\mu}\!_{\alpha}\,\epsilon^{\nu)\alpha\beta\gamma}\
a_{\beta}\ u_{\gamma}
+ h^{(\mu}\!_{\alpha}\,\epsilon^{\nu)\beta\gamma\delta}
\ \nabla_{\beta}H^{\alpha}\!_{\gamma}\ u_{\delta}\nonumber\\
& & + \ \sfrac{\k0}{2}\,f'\,^{-1}\,\left[\ h^{(\mu}\!_{\alpha}\,
h^{\nu)}\!_{\beta}\,\nabla^{\alpha}T^{\beta}\!_{\gamma}\,u^{\gamma}
- h^{(\mu}\!_{\alpha}\,h^{\nu)}\!_{\beta}\left(T^{\alpha\beta}
\right)\dot{ } + \sfrac{1}{3}\,\dot{T}\,h^{\mu\nu}\ \right]
\nonumber\\
& & + \ \sfrac{1}{2}\,f'\,^{-1}\,f''\,h^{(\mu}\!_{\alpha}\,
h^{\nu)}\!_{\beta}\,\left[\ \dot{R}\left(R^{\alpha\beta}
-\sfrac{1}{3}\,R\,g^{\alpha\beta}\right)
- \nabla^{\alpha}R\,R^{\beta}\!_{\gamma}\,u^{\gamma}\right.
\nonumber\\
& & +\ \left.\nabla^{\alpha}\left(\nabla_{\gamma}\nabla^{\beta}R
\right)\,u^{\gamma} - \left(\nabla^{\alpha}\nabla^{\beta}R
\right)\dot{ }\ \right],\\
\l{divh}
h^{\mu}\!_{\alpha}\,h^{\gamma}\!_{\beta}\,\nabla_{\gamma}
H^{\alpha\beta} & = &
- \,\epsilon^{\mu\alpha\beta\gamma}\ \sigma_{\alpha\delta}\
E^{\delta}\!_{\beta}\ u_{\gamma}
+ 3E^{\mu}\!_{\alpha}\,\omega^{\alpha}\nonumber\\
& & + \ \sfrac{\k0}{2}\,f'\,^{-1}
\,\epsilon^{\mu\alpha\beta\gamma}\,
\nabla_{\alpha}T^{\delta}\!_{\beta}
\,u_{\gamma}\,u_{\delta}\nonumber\\
& & +\ \sfrac{1}{2}\,f'\,^{-1}\,f''\,
\epsilon^{\mu\alpha\beta\gamma}\,
\left[\ \nabla_{\alpha}\nabla_{\beta}\nabla^{\delta}R
- \nabla_{\alpha}R\,R^{\delta}\!_{\beta}\ \right]\,u_{\gamma}\,
u_{\delta},\\
\l{hdot}
h^{\mu}\!_{\alpha}\,h^{\nu}\!_{\beta}\,\dot{H}^{\alpha\beta} & = &
3H^{(\mu}\!_{\alpha}\,\sigma^{\nu)\alpha} + H^{(\mu}\!_{\alpha}\,
\omega^{\nu)\alpha} - \Theta\,H^{\mu\nu}
- h^{\mu\nu}\,H_{\alpha\beta}\,\sigma^{\alpha\beta}\nonumber\\
& & + \ 2E^{(\mu}\!_{\alpha}\,\epsilon^{\nu)\alpha\beta\gamma}\
a_{\beta}\ u_{\gamma}
- h^{(\mu}\!_{\alpha}\,\epsilon^{\nu)\beta\gamma\delta}\
\nabla_{\beta}E^{\alpha}\!_{\gamma}\ u_{\delta}\nonumber\\
& & + \ \sfrac{\k0}{2}\,f'\,^{-1}\,h^{(\mu}\!_{\alpha}\,
\epsilon^{\nu)\beta\gamma\delta}\,\nabla_{\beta}
T^{\alpha}\!_{\gamma}\,u_{\delta}\nonumber\\
& & + \ \sfrac{1}{2}\,f'\,^{-1}\,f''\,h^{(\mu}\!_{\alpha}\,
\epsilon^{\nu)\beta\gamma\delta}\,\left[\ \nabla_{\beta}R\,
R^{\alpha}\!_{\gamma} - \nabla_{\beta}\nabla_{\gamma}
\nabla^{\alpha}R\ \right]\,u_{\delta}.
\eea
These equations generalise the analogous equations given for the
quadratic Lagrangian theories of gravity \c{maatay94} to the $f(R)$
case with a general energy-momentum tensor $T_{\mu\nu}$.
\\

And finally we derive, by use of the Gau\ss\ equation,
which holds for spacetime configurations with a
hypersurface-orthogonal matter fluid flow ($\omega_{\mu\nu}=0$),
the 3-Ricci curvature tensor of the spacelike 3-surfaces orthogonal
to $u^{\mu}$. The 3-Riemann curvature tensor
$^3R_{\mu\nu\rho\sigma}$ is given by
\be
^3R_{\mu\nu\rho\sigma} = \ ^3R_{\mu\rho}\,h_{\nu\sigma}
- \ ^3R_{\mu\sigma}\,h_{\nu\rho} + \ ^3R_{\nu\sigma}\,h_{\mu\rho}
- \ ^3R_{\nu\rho}\,h_{\mu\sigma}
- \sfrac{1}{2}\,^3R\left(h_{\mu\rho}\,h_{\nu\sigma}
- h_{\mu\sigma}\,h_{\nu\rho}\right)\ .
\ee
Using the field equations (\r{id4}), we can split the 3-Ricci
tensor $^{3}R_{\mu\nu}$ into its trace
\bea
^3R & = & 2\,\sigma^{2} - \sfrac{2}{3}\,\Theta^{2}\\
& & + \,f'\,^{-1}\left[\ f
+ f''\left(2\,h^{\mu\nu}\nabla_{\mu}\nabla_{\nu} R
- 3\,\Box R \right) + f'''\,(2\,\dot{R}^{2}-\nabla_{\mu}
\nabla^{\mu}R)
+ \k0\left(2\,T_{\mu\nu}u^{\mu}u^{\nu}+ T\right)\ \right]\ ,
\nonumber
\eea
and its trace-free part
\bea
^3R_{\mu\nu} - \sfrac{1}{3}\ ^3R\,h_{\mu\nu} & = &
h^{\alpha}\!_{(\mu}\,h^{\beta}\!_{\nu)}\left(\nabla_{\alpha}
a_{\beta} - \dot{\sigma}_{\alpha\beta}\right)
- \Theta\,\sigma_{\mu\nu} + a_{\mu}\,a_{\nu} - \sfrac{1}{3}\,
h_{\mu\nu}\,\nabla_{\rho}a^{\rho}\\
& & + \,f'\,^{-1}\left(h^{\alpha}\!_{\mu}\,h^{\beta}\!_{\nu}
- \sfrac{1}{3}\,h_{\mu\nu}\,h^{\alpha\beta}\right)\left[\ f''\,
\nabla_{\alpha}\nabla_{\beta}R + f'''\,\nabla_{\alpha}R\,
\nabla_{\beta}R + \k0\,T_{\alpha\beta}\ \right]\ .\nonumber
\eea
These relations then complete the generalisations of the
kinematical and the dynamical equations given by Ehlers and Ellis
\c{ehl61,ell71} for GR and those given by \c{maatay94} for the
quadratic gravity to the case of $f(R)$ theories with a general
energy-momentum tensor $T_{\mu\nu}$.
\section{FLRW Cosmological Models in $f(R)$ Theories}
In this section we look for conditions for the kinematic
characterisation of FLRW geometry in the context of $f(R)$
theories. To begin with let us recall that the kinematic
conditions, in the context of GR and for perfect fluid spacetimes,
are given by
\be
\l{a}
a_{\mu} = \omega_{\mu\nu} = \sigma_{\mu\nu} = 0\ ,
\ee
which with coordinates $x^{\alpha} = (t,x^{a})$
allow the metric to be
expressible in the following form
\be
\l{b}
ds^2 = -\,dt^2 + A(t)^2\,\Lambda_{ab}(x^{c})\,dx^{a}\,dx^{b}\ ,
\ee
such that the spacelike 3-surfaces defined by the metric
$\Lambda_{ab}(x^{c})$ have {\em constant} curvature.  In usual
Einstein's general relativity, the conditions (\r{a}) along with
the constraint equations (\r{id9}) - ({\r{id11}) imply the
vanishing of the Weyl tensor. However, for the $f(R)$ generalised
equations with perfect fluid matter source, which we shall assume
here, only $H_{\mu\nu}$ is immediately zero, and therefore to
show that $E_{\mu\nu} = 0$ we need to show that the Ricci scalar is
spatially homogeneous. To do this, we recall \c{anderson} that the
kinematic conditions (\r{a}) imply the existence of the metric
\be
\l{c}
ds^2 = -\,dt^2 + S(t,x^{c})^2\,\lambda_{ab}(x^{c})\,dx^{a}\,dx^{b}
\, ~~~~ u^{\mu} = {\delta^{\mu}}_0\ ,
\ee
where $\sfrac{1}{3}\,\Theta = \nabla_{0}S/S$. The $(0a)$ field
equations (\r{id3}) together with the constraint equations
(16) imply that
\be
\l{d}
\partial_{a}\partial_{0}\left(\ln S\right)
= \sfrac{1}{3}\,\nabla_{a}\Theta
= -\,\frac{1}{2\,f'}\,\nabla_{a}\nabla_{0}\left(f'\right)\ .
\ee
Now, if we assume that our matter fluid source obeys a barotropic
 equation of state of the form $p = p(\rho), ~~{dp(\rho)/d \rho}
\neq 0$, where $p$ and $\rho$ denote the isotropic pressure and the
total energy density of the matter fluid respectively, then the
conservation equations ${\nabla_{\nu}T^{\mu\nu}} = 0$ with the
kinematic conditions (\r{a}) imply that
\be
\l{e}
h^{\nu}\!_{\mu}\,\nabla_{\nu}p = h^{\nu}\!_{\mu}\nabla_{\nu}\rho
= h^{\nu}\!_{\mu}
\nabla_{\nu}\Theta = 0\ .
\ee
In the coordinates of (\r{c}), (\r{e}) gives $\Theta = \Theta(t)$,
therefore, (\r{d}) implies $S = A(t)\,B(x^{a})$, which reduces the
metric (\r{c}) to the form (\r{b}).

The next step is to show that $\Lambda_{ab}(x^{c})$ has constant
curvature.  Again, from (\r{d}) and (\r{e}) we have
$\nabla_{a}\nabla_{0}
\left(f'\right) = 0$ which, using the connection derived from
(\r{b}), integrates to
\be
\l{f}
f' = A(t)\,g(x^{a}) + B(t)\ ,
\ee
where $g$ and $B$ are arbitrary functions of $x^a$ and $t$
respectively. To proceed from here is not as straightforward
as the proof in \cite{maatay94}. We use equations (5) and (9)
to eliminate the term involving $f$. The resulting equation,
using (30), (31) and the perfect fluid form of the
energy-momentum tensor is
\be
{6\dot A^2\over{A^2}} - {6\ddot A\over{A}}
+ {^{3}{R^{*}}\over{A^2}}
= {1\over f'}\left[\ {1\over{A^2}}\,\Lambda^{ab}\,\nabla_{a}
\nabla_{b}\left(f'\right) + 3\,\nabla_{0}\nabla_{0}\left(f'
\right) + 3\k0\left(\rho+p\right)\ \right]\ ,
\ee
where $^{3}{R^{*}}(x^c)$ is the 3-Ricci scalar of the spacelike
3-surfaces defined by the metric $\Lambda_{ab}(x^{c})$.
Using equation (34) and the form of the metric (30), equation (35)
can be reduced to (provided $A\neq 0$)
\be
h(x^a) + B(t)\,^{3}{R^{*}}(x^a) + 9\dot A(t)^2\,g(x^a) + C(t) = 0
\ee
where $h(x^a)$ and $C(t)$ are given functions and $A$, $B$ and $g$
are defined by (30) and (34). It is clear from equation (36) that
$\nabla_a g = 0$ and consequently that
$\nabla_{a}\left(f'\right) = 0$, which in turn implies
\be
\l{h}
h^{\nu}\!_{\mu}\nabla_{\nu}R = 0\ .
\ee
Now to show that $\Lambda_{ab}(x^{c})$ has constant curvature we
proceed as follows. Equation (\r{h}) and the (GR) FLRW kinematic
conditions (\r{a}) imply \cite{anderson}
\be
\l{i}
\nabla_{\mu}\nabla_{\nu}R = \ddot{R}\,u_{\mu}\,u_{\nu}
- \left(\dot{R}\,\dot{A}/A\right)h_{\mu\nu}\ ,
\ee
which by (15) and (28) implies
\be
\l{j}
E_{\mu\nu} = 0 =~ ^{3}R_{\mu\nu} - \sfrac{1}{3}\,^{3}R\,h_{\mu\nu}
\ ,
\ee
where $^{3}R$ is the 3-Ricci tensor formed from $h_{ab} (t,x^c)$,
proving that the spacelike 3-surfaces $t = \mbox{constant}$ are
intrinsically isotropic. And finally we need to show that such
3-surfaces are isotropically embedded.  The reasoning for this
remains unchanged to the one given in \cite{maatay94} for the
quadratic theories.  First of all recall that \cite{stephani}
\be
\l{one}
R_{ab} = ~^{3}{{R}^*}_{ab}(x^{c}) + \left(A\,\ddot{A} + 2\,{\dot
A}^2\right)\Lambda_{ab}\ ,
\ee
where $ ^{3}{{R}^*}_{ab}(x^{c})$ is the 3-Ricci tensor formed from
$\Lambda_{ab}(x^{c})$. It can also be shown, using the field
equation (11) together with (9) and (5), that
\be
\l{two}
R_{ab} = \left(\sfrac{1}{3}\,A^2\,R - A\,\ddot{A}\right)
\Lambda_{ab}\ .
\ee
Now since $A$ and $R$ are functions of $t$ only, equations
(\r{one}) and (\r{two}) imply that $^{3}{{R}^*}_{ab}(x^{c}) =
2\,K\,\Lambda_{ab}(x^{c})$, where
\be
2\,K = \sfrac{1}{3}\,A^2\,R - 2\,A\,\ddot{A} - 2\,{\dot A}^2
= ~\mbox{constant}\ ,
\ee
and $K$ is the constant curvature of $\Lambda_{ab}(x^{c})$.  This
then shows that the spacelike 3-surfaces $t=\mbox{constant}$ are
isotropic and isotropically embedded and therefore the metric is
FLRW. The corresponding Bianchi identities (\r{dive}) - (\r{hdot})
are all identically satisfied, except for the $E_{\mu\nu}$
propagation equation (\r{edot}), which reduces to the field
equation (11).
\\

\noindent
In this way we have a generalisation of the result given in
\cite{maatay94}, thus:
\\

\noindent
{\it In the general $f(R)$ theories of gravity, a perfect fluid
spacetime with vanishing vorticity, shear and acceleration is FLRW
only if the fluid has in addition a barotropic equation of state.}
\\

An important consequence of this result is that the
Ehlers--Geren--Sachs theorem also holds for the $f(R)$ theories of
gravity.

In this sense then, the main conclusions of Maartens and Taylor
remain stable to any perturbations of the action. This represents
a significant factor in favour of the higher-order theories given
that the EGS theorem is one of the key results motivating the
standard big-bang model of the Universe.\\

Finally, it has recently been shown \c{sme95} that the observed
almost-isotropy of the cosmic microwave background radiation (CMBR)
implies that the Universe has been almost spatially homogeneous and
isotropic since the photon decoupling \footnote{This amounts to an
almost EGS theorem.}.  In terms of covariant perturbation theory
\c{sme95}, this means that deviations of the spacetime kinematics
and dynamics from FLRW are at most of order one with respect to
some smallness parameter $\epsilon$. In other words, if a quantity
disappears in a FLRW spacetime, it is of order one (or higher) in
an almost-FLRW spacetime, which allows the metric to be written in
a perturbed FLRW form \c{sme95}.  Subsequently, Taylor and Maartens
\c{tm95} have shown that this theorem also holds in quadratic
theories of gravity and may have applications to the very early
Universe.
\\

Now the results of the section 2, together with the above result
that EGS theorem also holds for the general $f(R)$ theories of
gravity, imply that the almost-EGS theorem shown to hold for
quadratic gravity of \c{tm95} is further generalisable to the case
of $f(R)$ theories. We see this as follows.

We use the formalism of \c{sme95,tm95} and elaborate only on the
deviations from the proof of \c{tm95} where necessary.  We consider
an acceleration-free ($a^{\mu} = 0$) expanding congruence ($\Theta
> 0$), where the kinematics is determined by the shear, vorticity
and expansion. It can be shown in a manner analogous to that in
\c{sme95}, that the components of the energy-momentum tensor, which
deviate from perfect fluid form, are not dissipative
quantities. Then the energy flux and the anisotropic stress,
$q_{\mu}$ and $\pi_{\mu\nu}$, measure the deviation of the
distribution function from isotropy and, since they vanish in FLRW
spacetimes, are first order quantities, i.e.
\be
\l{aa}
q_{\mu}~,~\pi_{\mu\nu} = O[1]\ .
\ee
{}~From the conservation equations it is possible to derive the
kinematic conditions (\r{a}) to first order, giving
\be
\l{ab}
\omega_{\mu\nu}~,~\sigma_{\mu\nu} = O[1]\ .
\ee
Furthermore, it is also possible to establish the conditions
(\r{e}) to first order:
\be
\l{ac}
h^{\nu}\!_{\mu}\,\nabla_{\nu}p~,~h^{\nu}\!_{\mu}\nabla_{\nu}\rho
{}~,~h^{\nu}\!_{\mu}\nabla_{\nu}\Theta = O[1]\ .
\ee
Equations (\r{aa}), (\r{ab}) and (\r{ac}) are crucial to proving
almost-homogeneity of the scalar curvature. From the constraint
equations (\r{id9}) we have that
\be
\l{ad}
h^{\mu}\!_{\nu}\left(\nabla^{\nu}\nabla_{\rho}f'\right)u^{\rho}
= O[1]\ .
\ee
By taking the spatial derivative of the generalised Raychaudhuri
equation (\r{id6}) and using (\r{ad}) we obtain
\be
\l{ae}
\left[\ f''\left(8\,\dot\Theta + 2\,\Theta^2 - 2\,R\right)
- f'\ \right]\,h^{\nu}\!_{\mu}\nabla_{\nu}R = O[1]\ ,
\ee
which in view of the fact that $R$,
$\Theta$ and $\dot\Theta$ are non-zero in the exactly isotropic (FLRW)
case, they would be $O[0]$ and we can deduce
\be
\l{af}
h^{\nu}\!_{\mu}\nabla_{\nu}R = O[1]\ .
\ee
Then from (\r{id8}) and (\r{id11}) we obtain

\be
\l{ag}
E_{\mu\nu}~,~H_{\mu\nu} = O[1],
\ee
which establishes the almost-FLRW kinematics to $O[1]$. Equation
(\r{af}) implies also that the 3-surfaces of constant energy
density are almost-isotropic and almost-isotropically embedded. We
see this from equations (28) and (42), where it follows from
(\r{af}) that
\be
^3R_{\mu\nu} - \sfrac{1}{3}\ ^3R\,h_{\mu\nu} = O[1]
\ee
and
\be
2\,K = \sfrac{1}{3}\,A^2\,R - 2\,A\,\ddot{A} - 2\,{\dot A}^2
{}~~,~~\dot K = O[1].
\ee
Thus, it is possible to recover all the standard relations
governing an FLRW model of the Universe to $O[0]$, and in
particular, to show that there exists an almost-FLRW metric in
general $f(R)$ gravity. This then shows that the ``almost'' EGS
generalises to the case of $f(R)$ theories.
\section{Conclusion}
We have derived the generalised kinematic and dynamic equations
governing relativistic fluid dynamics (given by Ehlers and Ellis
for GR and for the quadratic theories by Maartens and Taylor) in
the case of generalised $f(R)$ theories of gravity and with a
general $T_{\mu\nu}$.  Such generalisation is of potential
importance as it gives an alternative framework in order to derive
results analogous to those obtained in GR for the generalised
theories of gravity.
\\

As an example of application of these results, we have generalised
the results of Maartens and Taylor to show that within the
framework of $f(R)$ theories a perfect fluid spacetime with
vanishing vorticity, shear and acceleration is FLRW only if the
fluid has in addition a barotropic equation of state. It then
follows that the Ehlers--Geren--Sachs theorem holds for general
$f(R)$ theories as well. In this sense, then, the results of
Maartens and Taylor are stable to perturbations of any order in the
action. We also show that the ``almost'' EGS theorem of Stoeger,
Maartens and Ellis and its generalisation derived by Taylor and
Maartens in the case of quadratic gravity also remains stable to
perturbations to the Lagrangian of the form given in equation (3).
\section*{Acknowledgements}
We would like to thank Roy Maartens for many helpful discussions.
SR is supported by a PPARC studentship, HvE is supported by a
grant from the Drapers' Society at QMW and RT is supported by
the PPARC UK, under grant No. H09454. DT thanks David Matravers
and Portsmouth University for hospitality and a research grant
and Astronomy Unit, QMW for visiting support.
\newpage


\end{document}